\documentclass[preprint,showpacs,preprintnumbers,amsmath,prl,amssymb,floats]{revtex4}
\usepackage[dvips]{graphicx}
\usepackage[english]{babel}
\usepackage{amsmath}
\usepackage{amssymb}
\usepackage{color}

\usepackage{amssymb,graphicx,color,amsmath,xfrac,bm,ifthen,lineno}
								\usepackage{oldstyle}
								\usepackage[mathletters]{ucs}
								\usepackage[utf8x]{inputenc}
								\usepackage{caption}
								\definecolor{lightblue}{rgb}{0.17,0.39,1}
								\definecolor{lightgreen}{rgb}{0.67,0.81,0.08}
								\definecolor{lightred}{rgb}{1,0.05,0.52}
								\usepackage[
								bookmarks,
								colorlinks=false, 
								breaklinks,
								pdftitle={Chiral}, 
								pdfauthor={C.M.Varma}
								]{hyperref}
								\hypersetup{
								linkcolor=black,
								citecolor=black,
								filecolor=black,
								urlcolor=black
								}

								\newcommand{\para}[1]{\left(#1\right)}

\newcommand{\be}{\begin{eqnarray}}
\newcommand{\ee}{\end{eqnarray}}

\begin{document}

\title{Chiral spin-order in some \\ purported Kitaev spin-liquid compounds}

\author{K.~A.~Modic}\affiliation{Max-Planck-Institute for Chemical Physics of Solids, Noethnitzer Strasse 40, D-01187, Dresden, Germany}
\author{B.~J.~Ramshaw}\affiliation{Laboratory of Atomic and Solid State Physics, Cornell University, Ithaca, NY 14853, USA}
\author{A.~Shekhter}\affiliation{National High Magnetic Field Laboratory, Florida State University, Tallahassee, FL 32310, USA}
\author{C. M. Varma}
\thanks{Visiting scientist}
\affiliation{Department of Physics, University of California, Berkeley, CA. 92740, USA}

\date\today
\begin{abstract}
We examine recent magnetic torque measurements in two compounds, $γ$-Li$_2$IrO$_3$ and RuCl$_3$, which have been discussed as possible realizations of the Kitaev model. The analysis of the reported discontinuity in torque, as an external magnetic field is rotated across the $c-$axis in both crystals, suggests that they have a translationally-invariant chiral spin-order of the from $<{\bf S}_i \cdot \big({\bf S}_j~ \times~ {\bf S}_k\big)> \ne 0$ in the ground state and persisting over a very wide range of magnetic field and temperature.  An extra-ordinary $|B|B^2$  dependence of the torque for small fields, beside the usual $B^2$ part,  is predicted due to the chiral spin-order. Data for small fields is available for  $γ$-Li$_2$IrO$_3$ and is found to be consistent with the prediction upon further analysis. Other experiments such as inelastic scattering and thermal Hall effect and several questions raised by the discovery of chiral spin-order, including its topological consequences are discussed. 
\end{abstract}
\maketitle

\section{Introduction}
In the last few decades, there has been much discussion of the possibility of  insulators with magnetic ions which do not order down to the lowest temperatures due to quantum fluctuations \cite{Savary-Balents-rev}. Such states have been given the name spin-liquids. The interest in such problems is high in view of their possible connection to emergent quantum numbers, fractionalization of excitations, etc.  
The theory, calculations and experimental realizations have been clear in one dimension. In two dimensions,  Kitaev \cite{Kitaev2006} has provided exact results on some models, while there have been many approximate discussions on several related models. The models are rather special and not easily realizable although impressive crystal symmetry analysis \cite{Jackeli2009} has led to the search for materials with the requisite anisotropic exchange. The fact that a variety of such compounds show no customary magnetic order down to temperatures an order of magnitude or more below their magnetic interaction energies, and are not spin-glasses, speaks for quantum fluctuations in a general way. But specific experimental signatures have been murky.  

We analyze clear and anomalous results from magnetic torque measurements in two compounds, $γ$-Li$_2$IrO$_3$ \cite{Modic, Modic1} and RuCl$_3$, which due to their structure and quantum-chemistry may host Kitaev-like exchange anisotropy between effective $S=1/2$ ions on hexagonal networks \cite{Jackeli2009} together with additional interactions. These compounds exhibit antiferromagnetic (AFM) order \cite{Biffin, Cao} at low temperatures and small applied magnetic fields. But thermal transport \cite{TT}, inelastic neutron scattering experiments \cite{INS}, and Raman spectroscopy \cite{Raman} suggest unusual properties in and outside of the AFM region that are not to be expected in AFMs. Suggestions have been made that these properties are characteristic of  Kitaev spin-liquids  \cite{Raman, itamar, Kim, Surendran}. We show here that the experimental results are consistent with a specific local order parameter, which  does however have topological properties.

\section{Magnetic Torque}
 A torque ${\bf τ}$ is generated when a magnetic field ${\bf B}$ is applied to an anisotropic magnetic crystal in a direction which is not 
one of the principal axes for the magnetic susceptibility ${\bf χ}$ \cite{LLPEDCM}.
\be
\label{tau}
{\bf τ} &=& {\bf M}({\bf B})× {\bf B},~~ {\bf M}({\bf B}) = - \frac{ d F}{d {\bf B}}.
\ee
$F$ is the free-energy and ${\bf M}_i$ is the magnetization in the $i-{th}$ direction. In the linear regime where, 
\be
{M}_i &=&  {\bf χ}_{ij} B_j,
\ee 
the torque in a magnetic material 
with orthorhombic or hexagonal symmetry normally follows the angle dependence
\be
\label{ordtau}
{\bf τ}_n(θ) =\frac{1}{2}(χ_{p}-χ_{c})\sin(2θ) B^2,
\ee
where $\theta$ is the direction of the magnetic field measured from the $a$-$b$ plane.
$χ_{c}$ and $χ_{p}$ are the susceptibilities with field in the $c-$axis and in one of the symmetry axes orthogonal to it, respectively. The above is true in a paramagnet or in an ordered AFM compound, however complicated the order may be, provided the AFM order preserves the principal axes invoked above. For larger $B$, the dependence on the torque only has even powers of $B$. We will briefly mention the angle dependence of the torque near the region close to the transition from the AFM to the paramagnetic phase due to a magnetic field later.

\begin{figure}
\includegraphics[width=0.8\textwidth]{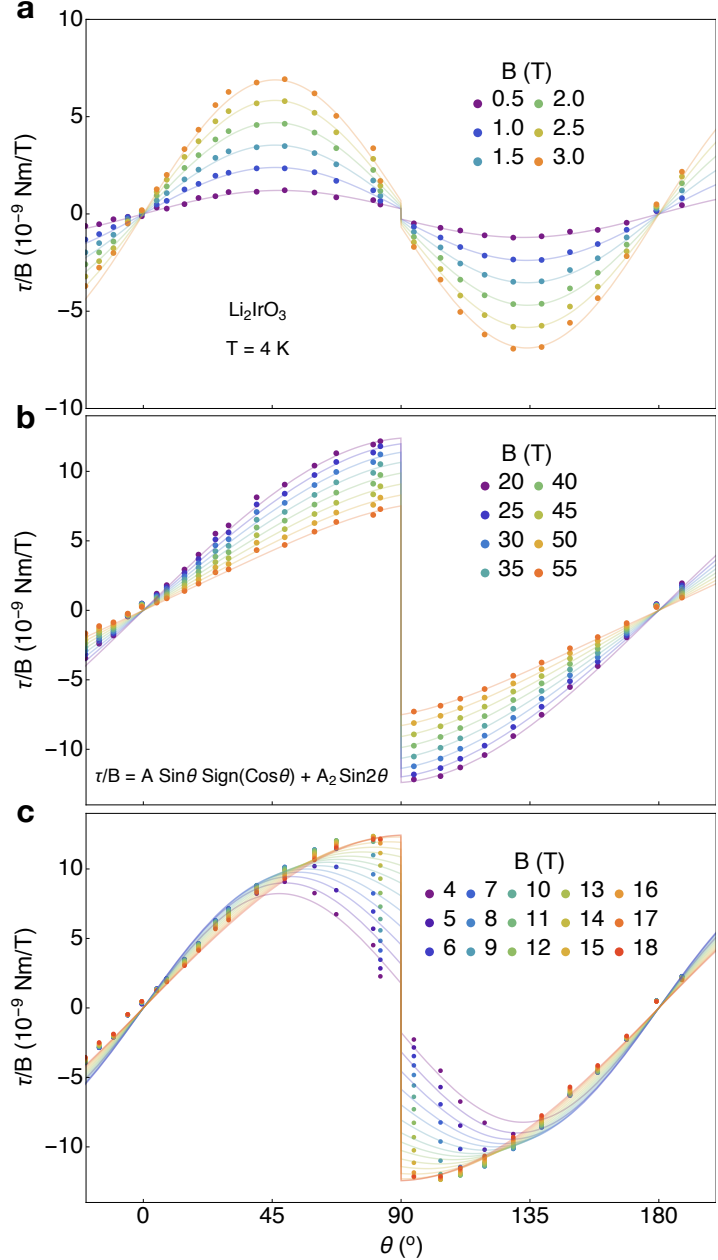}
\caption{Magnetic field evolution of the angle-dependent torque at low temperatures in  $γ$-Li$_2$IrO$_3$. The dots in the three panels give the angle dependence in three different field regions. In (a) the low field region in which the  AFM order is preserved for any angle of the applied field.  (b) shows the highest field region in which the AFM order is absent for field at any angle. (c) 
shows the intermediate field region in which the AFM order is suppressed above a field $H^*(\theta)$. The function $H^*(\theta)$ is shown in Fig. 2 of Ref. (\cite{Modic1}). The solid curves in all three panels are best fits to the data with the functional form given in panel (b). The field dependence of the coefficients A and A$_2$ are given in Fig. (\ref{Fig: a,a2}). The discrepancies of the fit in the intermediate field region are discussed in the text.
The data as a function of magnetic field at fixed angles has been shown in Ref. (\cite{Modic, Modic1}).}
\label{Fig:torque-iridate}
\end{figure}

\begin{figure}
\includegraphics[width=0.8\textwidth]{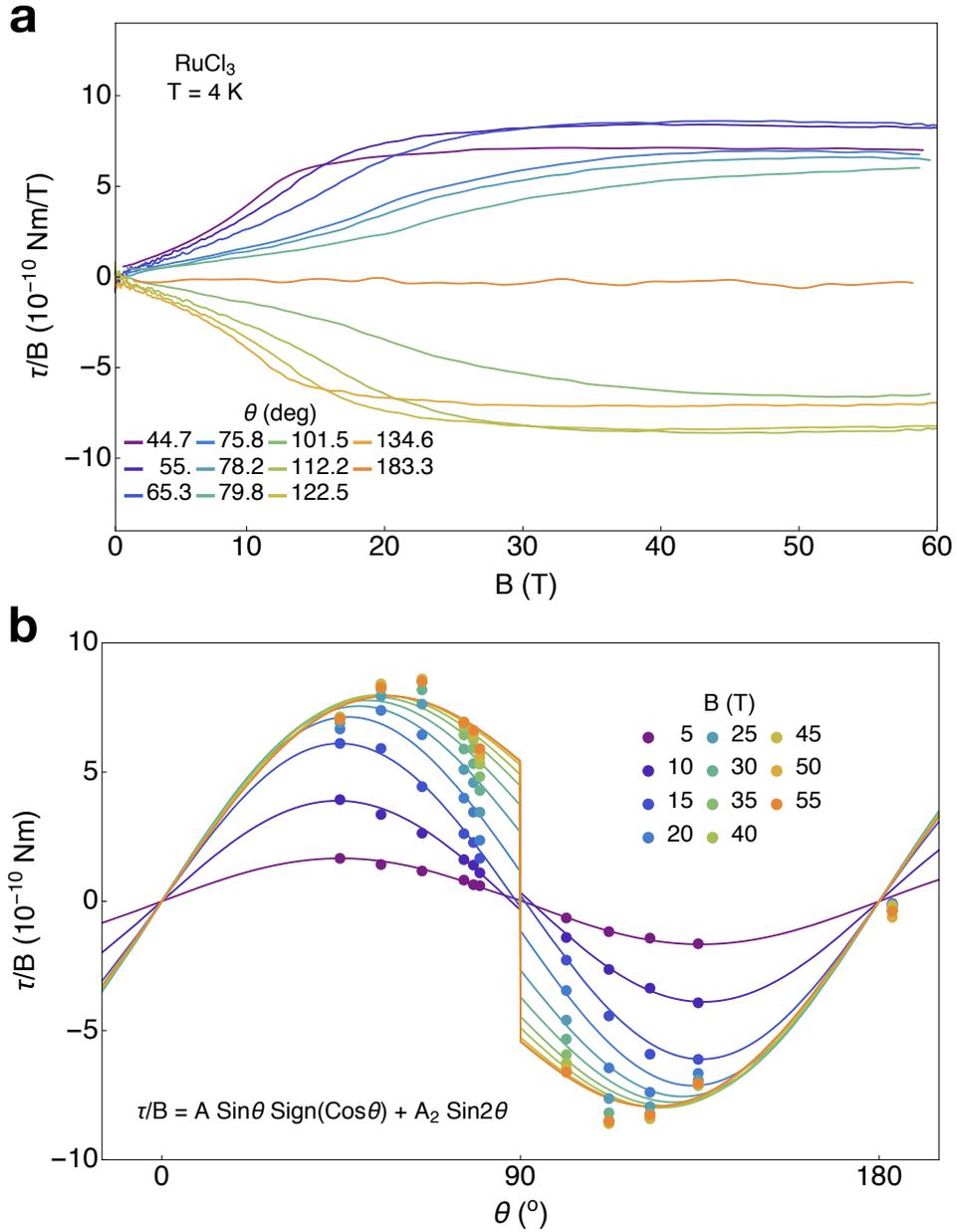}
\caption{Magnetic field evolution of the angle-dependent torque at low temperatures in RuCl$_3$. The first panel (a) shows the field dependence at various fixed angles and the second panel (b) shows the angle dependence at various fixed fields. T  More comprehensive results in RuCl$_3$, including for the discontinuity at near $\theta = \pi/2$, have been obtained recently, through measurements of  the magnetotropic coefficient \cite{Modic3, Modic2}. and are consistent with the results shown here.}
\label{Fig:torque-rucl3}
\end{figure}

The results for the torque measurements as a function of angle for various applied fields are shown in Fig. (\ref{Fig:torque-iridate}) for  $γ$-Li$_2$IrO$_3$ and in Fig. (\ref{Fig:torque-rucl3}) for RuCl$_3$. The available data for RuCl$_3$ is not as extensive as for $γ$-Li$_2$IrO$_3$. \cite{footnote}.  The data for $γ$-Li$_2$IrO$_3$ is shown separately in three different panels for three different field regions described in the figure caption.  At small enough fields, the results in both compounds are dominated by the angle dependence of Eq. (\ref{ordtau}) \cite{Modic}. At larger fields $ B ≳ B^*(θ, T)$, the {\it dominant} term in the angle-dependent torque has the anomalous angle dependence
\be
\label{extordtau}
\frac{{\bf τ}_a(θ)}{|B|} &=& |{\bf N}(B)| \sinθ \;\text{sign}\para{\cosθ}.
\ee
${\bf τ}(θ)/|B|$ jumps from its maximum positive value at $θ \approx (π/2)^-$ to its maximum negative value at $θ \approx (π/2)^+$ \cite{Modic1}. $\tau_a$ remains the same for $B \to -B$.  ${\bf N}(B)$ reaches a maximum at about 30 Tesla at  $T= 4$ K in  $γ$-Li$_2$IrO$_3$ and then slowly decreases with increasing field (Figure \ref{Fig:torque-iridate}). This slow decrease is consistent with the exchange interaction energy scale $J$, determined by the deviation from the Curie law at 200 K \cite{Modic}.  {\it In fact, as further discussed below, closer examination reveals that data at lower fields is also consistent with a torque which is the sum of the two terms with angular dependence of the form (\ref{ordtau}) and (\ref{extordtau})}. This behavior continues at temperatures and magnetic fields well beyond the AFM state \cite{Modic1}.

In RuCl$_3$, the discontinuity occurs as magnetic field crosses the direction perpendicular to the  honeycomb plane, suggesting that ${\bf N}(B)$ in Eq. \ref{extordtau} lies within the honeycomb plane. Furthermore, the discontinuity appears consistent with a six-fold modulation as the rotation plane of the magnetic field changes with the azimuthal angle $\phi$ \cite{Modic2}.  In $γ$-Li$_2$IrO$_3$, there are two inequivalent honeycomb planes which share the $c-$direction and are oriented azimuthally at $\sim \pm 35^\circ$ from the $b-$axis. In this system, the discontinuity in torque also occurs as magnetic field crosses the $c-$axis, which reflects the discontinuous behavior of the \emph{total} ${\bf N}(B) ={\bf N}_1(B)+{\bf N}_2(B)$, where ${\bf N}_1(B)$ and ${\bf N}_2(B)$ refer to the two inequivalent honeycomb planes.  

More comprehensive results than shown in (\ref{Fig:torque-rucl3}) for RuCl$_3$ have been obtained recently in the magnetotropic coefficient \cite{Modic3, Modic2}, which measures the angular derivative of torque. The discontinuity in torque manifests itself as a sharp peak in the magnetotropic coefficient as the field is moved across the c-axis.

As noted above, the angular dependence in (\ref{extordtau}) preserves the point-group symmetries of the crystal. It is the discontinuity when the field is turned across $\theta = \pi/2$ and $3\pi/2$  which is anomalous. The magnitude of the discontinuity depends on $B$. One might think that ${\bf N}(B)$ is an ordinary magnetization-vector which at high fields lies purely in the hexagonal planes and jumps as the angle of the field is changed across the c-axis. However, we have not found any spin-reorientation free-energy for a collinear or a non-collinear magnetic order parameter characterized by a vector at zero or non-zero ${\bf Q}$ or any two-dimensional magnetic tensor order parameter which gives the observed jump. 

What about the torque when there is a phase transition as a function of ${\bf B}$? There is an angle-dependent field $B^*(\theta)$ in both compounds at which there is a second order transition from the AFM phase to the paramagnetic phase. As $B \to B^*(\theta)$ (or fields where there is a transition from one AFM phase to another) there must be a rapid variation in the torque as a function of angle related to $dB^*/d\theta$ and so a departure from the $\sin (2 \theta)$ dependence of Eq. (\ref{ordtau}). But the functional dependence of the variation with angle in this case is field dependent unlike the dependence in Eq. (\ref{extordtau}) where only the amplitude depends on the field but the angle dependence is independent of it. Detailed calculations consistent with such an idea have been carried out \cite{valenti}. Their unimportance to the anomalies on which we have focussed here is seen in comparing Fig. (\ref{Fig:torque-iridate}) for $γ$-Li$_2$IrO$_3$, which follows Eq. (\ref{extordtau}), with Fig. (2c,d) in Ref. \cite{valenti}. Such an effect is irrelevant in the very high field region shown in (\ref{Fig:torque-iridate}-c and a) which are above and below $B^*$ for all $\theta$ in which very good fits are obtained to  Eq. (\ref{extordtau}).  The small but noticeable discrepancies to the fit to Eq. (\ref{extordtau}) in the intermediate region shown in (b) of the same figure may be ascribed to such effects. The available data in RuCl$_3$ is at present not extensive enough to quantitatively establish the relative magnitude of the effects.

Actually, an unambiguous signature of a new and interesting effect in $γ$-Li$_2$IrO$_3$ is provided by the prediction and observation of the B-dependence of ${\bf N}(B)$ at small $B$ which is discussed below. Similar low field data is needed  to ascertain the issue in RuCl$_3$.

\section{A hypothesis and its test}

 Consider the scalar operator formed of the solid angle subtended by three spins,  
 \be
\label{top}
{Ω} ≡ \frac{1}{2N}∑_{(i i' i"), \Delta =1,2} {\bf S}_{i \Delta}. \big({\bf S}_{i', \Delta} × {\bf S}_{i", \Delta} \big).
\ee
$\Delta$ labels the two triangular sub-lattices of a hexagonal unit-cell, labelled by $i$; for a given $\Delta$, $(i, i', i")$ label the three sites in the sub-lattice in a unit-cell {\it in an ordered way}, say clockwise with respect to the axis perpendicular to the hexagon. $N$ is the number of unit-cells, and 2 is the number of sub-lattices.  We find that the  simplest state which gives the observed properties is a state with a finite thermodynamic average $< \Omega>$.
We need consider only the case that $\Delta = 1,2$ contribute equally to $<\Omega>$. So, henceforth we will drop the subscript $\Delta$ as well as the factor $1/2$, with the understanding that $(i, i', i")$ refer to sites in the same sub-lattice in a unit-cell. (\ref{top}) can be easily generalized to more than one hexagonal plaquette per unit-cell. The order parameter $<{Ω}>$ is a scalar which odd under both time-reversal and all reflections - it is chiral. The product of time-reversal and chirality is preserved, as is translation by lattice vectors. 
%$Ω$ is hermitian, so the order parameter is real. 
Further it is stipulated that individual spins and pairs fluctuate so that $<{\bf S}_{i}> =0$ and  $<{\bf S}_{i'} × {\bf S}_{i"}> =0$, while the  thermodynamic average  $<{Ω}>$ maintains its fixed value. Such an order cannot be discovered by polarized neutron scattering. Other methods which may show consistency with such an order are discussed below.

Long ago, Herring \cite{Herring} derived that $i~ {\bf S_i. (S_j \times S_k)}$ appears in the permutation operator or the ring-exchange Hamiltonian for three-spins at sites $(i, j,k)$ in a magnetic insulator. 
A variational ground state wave-function proposed by Kalmeyer and Laughlin \cite{KL} (see also \cite{KR}) for spins in an insulator on a triangular lattice, as an alternative possibility to AFM order, has the symmetries of the order parameter $\Omega$. Wen, Wilczek and Zee \cite{WWZ}   discussed the order parameter $\Omega$ in the context of their description of anyonic excitations. Such an order parameter, which is equivalent to spin-currents within each unit-cell in the lattice may be derived to be locally stable in a mean-field theory from physically relevant interaction terms in the Hamiltonian analogously to the loop charge-currents in Ref. (\cite{cmv1, cmv2}).

A term proportional to the operator $\Omega^2$ is always allowed in the Hamiltonian. Given that there is a Hamiltonian for which the order parameter $<\Omega> \ne 0$, it follows, since  ${\bf S_i. (S_{i'} \times S_{i"})}$ is hermitian, that a term proportional to $<\Omega> \Omega$ belongs in the  Hamiltonian. Since a magnetic field ${\bf B}$ has the same symmetries as ${\bf S}$, a lowest order in ${\bf B}$ term found in the Hamiltonian \cite{footnote2} is
\be
\label{H'}
H' =  γ {\bf B}. ∑_{(i' i")}<{\bf S}_{i'} × {\bf S}_{i"}>(B)
\ee
 $γ$ is a coefficient  proportional to $<\Omega>$ and so formally includes in it the product of the eigenvalues of the parity and time-reversal operators with the product remaining invariant.  $(i', i")$ are also ordered in a specific way following the definition after Eq. (\ref{top}). The observed behavior in Eq.~(\ref{extordtau}) can be understood if 
\be
{\bf N}(B) = γ ∑_{(i' i")}<{\bf S}_{i'} × {\bf S}_{i"}>(B).
\ee
${\bf N}(B)$ is even under time-reversal and odd under parity, and ${\bf N}(0) =0$, as stated above.  It may be called a {\it quantum screw vector} because it is characterized by its helicity and magnitude. In considering the contribution of $(\ref{H'})$ to the ground state energy, we take ${\bf N}(B)$ to lie in the hexagonal planes for all $B$'s under consideration due to anisotropies in the microscopic Hamiltonian, the details of which are not known. The dot product in Eq. (\ref{H'}) includes both the geometric angle between ${\bf B}$ and ${\bf N}({\bf B})$ as well as the product of the helicity of these two vectors. Beside the angle-dependence between the vectors ${\bf B}$ and ${\bf N}$, we must therefore also take into account that the helicity of ${\bf N}$ is picked by the helicity of the projection of ${\bf B}$ on ${\bf N}$. 
%This is because for ${\Omega}$ to have a fixed value while $\av{{\bf S}_i} = \av{{\bf S}_{i'} × {\bf S}_{i"}} = 0$, the direction of  $({\bf S}_{i'} × {\bf S}_{i"})$ must flip on flipping the projection of ${\bf S}_i$ on $({\bf S}_{i'} × {\bf S}_{i"})$. The direction of ${\bf S}_i$ is itself determined by the direction of ${\bf B}$.% 
Therefore the change of the ground state energy on applying a field ${\bf B}$ has a contribution,

%Torque is generated by ``rotating" the screw with a magnetic field, as distinct from the classic result (\ref{tau}), where ${\bf M}$ acts as a lever arm turned by $B$. Then the angular dependence of the form (\ref{extordtau}) follows because the sign of the screw helicity is picked by projection of ${\bf B}$ on to $\av{{\bf S}_{i} × {\bf S}_{i'}}$. Another way of seeing this is that for ${\Omega}$ to have a fixed value while $\av{{\bf S}_i} = \av{{\bf S}_{i'} × {\bf S}_{i"}} = 0$, the direction of  $({\bf S}_{i'} × {\bf S}_{i"})$ must flip on flipping the direction of ${\bf S}_i$. The direction of ${\bf S}_i$ is itself determined by the direction of ${\bf B}$. More precisely and summarizing, the free energy on applying a field ${\bf B}$ is changed by,%
\be
\label{BN}
δ E_a(B) = -\gamma |B| |{\bf N}(B)||\cos θ| f(\phi). 
 \ee
%In \ref{BN}, $\circ$ implies the dot-product of the direction of ${\bf B}$ with that of ${\bf N}$, as well as the product of their helicities. The former as well as the latter change the sign of $\cos(\theta)$ on ${\bf B}$ going across $\pi/2$.   Hence $|\cos \theta|$ in the second equality in (\ref{BN}).
\noindent
Obviously the direction of ${\bf N}({B})$ in the plane is set by the  direction of  of ${\bf B}$ projected to the plane. $f(\phi)$ is the dependence on the azimuthal angle of the magnetic field. It should respect the reflection symmetry of the planes passing through the c-axis. Details of the $f(\phi)$ depend on the quantization axis for the spins, which are determined by the microscopic Hamiltonian and are in general different for different sites. We cannot say more about this without knowledge of the microscopic Hamiltonian. 

The sign of $\gamma$ is picked to be positive to give the state with the lower value of energy. The anomalous contribution to the torque $\tau_a(B)$ derived from the ground state energy $δ E_a(B)$, using Eq. (\ref{tau}) is,
\be
\tau_a(B) = \frac{d~ δ E_a(B)}{d \theta}.
\ee
 $\tau_a(B)$ has precisely the form  (\ref{extordtau}) with which experiments have been fitted; it changes sign across $\cos(θ) = 0$ and its magnitude is proportional to $\sin(\theta)$. It is invariant under ${\bf B} \to -{\bf B}$ and also preserves all the point group symmetries as in the experiments.  
 
 \begin{figure}
\includegraphics[width=1.0\textwidth]{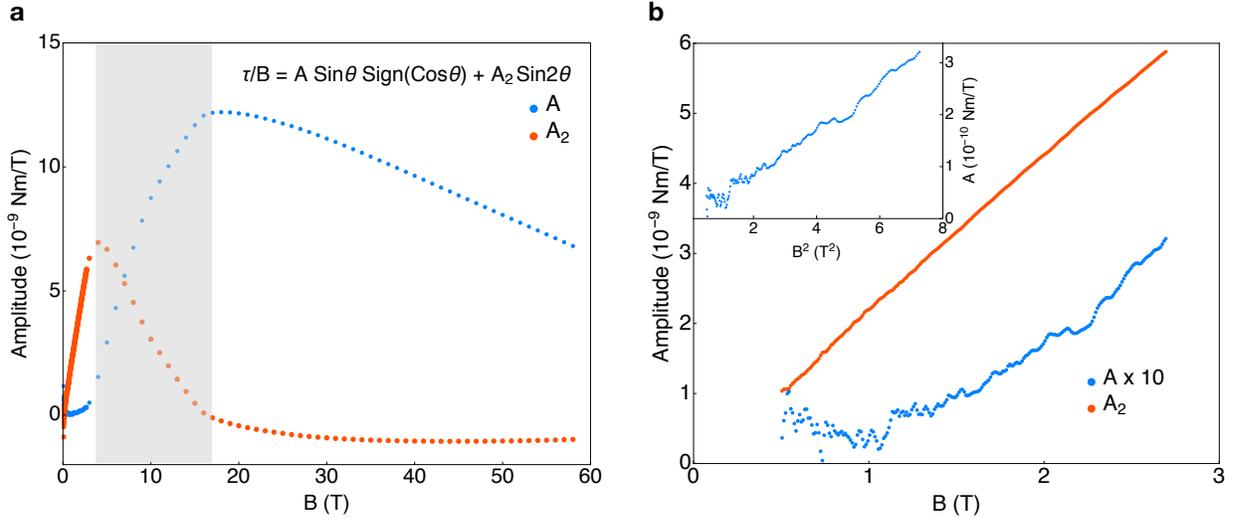}
\caption{(a) The coefficients $A$ and $A_2$, determined by fitting the  angle-dependence of the $\tau/B$ in Figure (\ref{Fig:torque-iridate}) and more such data at fixed temperature to $A~\text{sign}(\text{cos}θ)\text~{sin}θ$+$A_2~\text{sin}2θ$, as a function of magnetic field.  (b) The low-field dependence of $A$ and $A_2$. $A$ is multiplied by 10 for viewing on the $A_2$ scale. $A$ plotted against $B^2$ (inset) displays the $B^3$ dependence of the anomalous component of the torque at low fields. The shaded region in the first figure shows the region in which AFM is found below $B < B^*(\theta)$, the latter varies from 3 tesla for field in the hexagonal planes to 18 tesla for field normal to them Ref. (\cite{Modic}).}
\label{Fig: a,a2}
\end{figure}

There can be no linear (or odd power) dependence of $|{\bf N}|(B)$ on $B$. A prediction which follows is that the leading dependence of $|{\bf N}(B)| ∝ B^2$, i.e. $\tau_a$ proportional to $|B| B^2$. As discussed above, this follows from the symmetries of the chiral spin-order and the fact that it involves three spin-operators, each of which is tuned by ${\bf B}$. More specifically, given the spin-structure of ${\bf N}$, $\tau_a(\theta, \phi)$ depends on the product of the two orthogonal components of the field in the hexagonal planes with direction determined by the appropriate quantization axes, and of the component perpendicular to the plane which is a natural quantization axis. One may therefore also understand the observed discontinuity in the torque as follows: When the component of the field perpendicular to the plane and one of the components in the plane is held fixed and the other component in the plane changes sign, the torque must also change sign. This obviously happens when $\theta$ is turned across $\pi/2$. 

 For $\phi$ in a symmetry direction, the predicted field dependence of the anomalous torque at low fields has been tested by a detailed analysis of the data for the iridate compound which is given in Fig. (1a) of Ref. \cite{Modic}. This shows $\tau/|B|$ continuously as a function of $B$ at multiple field orientation angles. More than a hundred field slices are taken from this data and the angle dependence at these fixed fields is then fit to the sum of the two terms (\ref{ordtau}) and \ref{extordtau}, with denser field slices at low field to ascertain the dependence on magnetic field. The results are shown in Fig. (\ref{Fig: a,a2}).  $A_2(B)$ is the coefficient of the term proportional to $\sin(2 \theta)$, i.e. the normal term proportional to the anisotropy of the magnetic susceptibility. $A(B)$ is the coefficient of the term $\sin(\theta){\text{sign}}(\cos(\theta))$, i.e. it is proportional to $|{\bf N}(B)|$. The low field results are shown in an expanded form in Fig. (\ref{Fig: a,a2}-b), and show the predicted $B^2$ dependence up to 3 tesla. This is the maximum field where the low-field angle dependence can be fit without crossing the angle-dependent AFM phase boundary. The high-field behaviour of $A(B)$ and $A_2(B)$ can only be extracted above 18 tesla, outside of the shaded region in figure (\ref{Fig: a,a2}-a). A magnetoresistive contribution inherent to the torque detection method is removed by anti-symmetrization of the data. A zero-field offset due to the bridge circuit used in the torque measurement is removed such that torque is zero at zero field. However, whether these systematic effects are removed or not, the qualitative behavior of the $A$'s remains the same. We also note that in the low-field limit, $A_2$ dominates the total torque signal and we suggest direct measurements  of $M(B)$ to support that the leading order correction goes as $B^2$.

Just as AFM order does not give the observed jump of $\tau/B$ as a function of angle of $B$, it does not give a $\tau/B \propto B^2$ at low fields at any angle. In fact, for an AFM order, the free-energy must contain only even powers of $B$, therefore $\tau/B$ contains only odd powers of $B$. Similarly, the deviations from $\sin(2\theta)$ discussed in Ref. (\cite{valenti}) give $\tau/B$ with only odd powers of $B$.

 Low field torque data is not available for RuCl$_3$. Similar behavior as in Fig. (\ref{Fig: a,a2}) in this compound would be an unambiguous proof of chiral order in that compound.

In the experiments $|{\bf N}(B)|$ has a broad peak at an intermediate field and then decreases very slowly.
The slow decrease of this component at larger $B$ indicates decay of the chiral order parameter at an energy scale of the large bare magnetic couplings in the compound indicated by the Weiss-constant. 

If $|{\bf N}(B)|/B^2 \ne 0$, it follows that the order parameter ${Ω} \ne 0$ in zero applied field. It co-exists with the AFM order parameter ${\bf{M}}({\bf Q})$, and continues at temperatures and fields beyond where ${\bf{M}}({\bf Q}) = $ 0. 

We have found that there is a  steep rise in the coefficient $A$ near $B = B^*(\theta)$ where the AFM order and the coefficient A$_2$ begins to sharply decrease. This is consistent with an allowed coupling of the form
proportional to $u |{\bf{M}}({\bf Q})|^2 |{\Omega}|^2$, where $u$ is a repulsive coupling energy.  
 \subsection{Relation to Kitaev states} 

The ground state of the Kitaev model preserves time-reversal invariance unlike ${Ω} \ne 0$. (See however Ref(\cite{kivel}) for Kitaev model on a decorated honeycombe lattice.) An external magnetic field has no effect on the ground state (or the excitations) in the Kitaev model to order $B$ or $B^2$. For a magnetic field coupling as $∑_{α = x,y,z} B_{α} S_{α}$, a state with $Ω$ is generated to $O(B_xB_yB_z/J^3)$, where $J$ are the three couplings in the model assumed equal \cite{Kitaev2006}. In effect at this order the flux $w(p) = ∏_{i\subset p} S_i$ around the hexagonal plaquettes $p$ in the Kitaev model, which has a finite expectation value in the ground state in the absence of the field, breaks up into a sum of the expectation values $<Ω>$ of the two sub-lattices.  In the Kitaev model, an anomalous torque related to ${\bf N}({\bf B})$ is also expected with a discontinuity near $θ = π/2$, but such a torque would be proportional to $B^6$ at low fields as opposed to $O(B^3)$ observed in the experiments discussed above. 

\section{Some properties of the Chiral spin-ordered state}

Since $<Ω>$ breaks time-reversal, an internal magnetic field is generated. It may be observed by Kerr effect and by muon resonance. Similarly, breaking of chirality should be visible in second harmonic generation and in optical polarimetry. 

There have been several inelastic scattering experiments - neutron scattering \cite{INS} and Raman \cite{Raman} seeing a continuum of excitations carrying angular momenta of $± 1$. 
Continua of excitations are not to be expected in a spin-wave theory for conventional ordered states, especially at long wavelengths.  In a state with a ground state expectation value $Ω$, the {\it simplest} excitations  for a given total momentum ${\bf q}$ are expected to form a continuum. This is because the simplest low-energy excitation, formed from linear combinations of local $± 1$ excitations of ${\bf S}_i$ must be accompanied by excitations of $({\bf S}_j × {\bf S}_k)$ to correspond to the lowest local change in $Ω$. As discussed below, one should also expect  topological excitations. 

Given the expectation value $<{Ω}>$, thermal Hall conductivity $κ_{xy}$ is to be expected because of chiral surface states accompanying such an order parameter. Kitaev predicts a quantized value for this quantity in a (large) field due to field induced chiral spin-order when the bulk ground state has a gap \cite{Kitaev2006}. 
In RuCl$_3$, there is indeed good evidence for a finite $κ_{xy}$ \cite{TT}. Its value is even quantized to the predicted value but only in an intermediate field regime. The field dependence both at lower and at higher fields is complicated \cite{Balents, Moore} and further theory and experiments are required to understand it. Such measurements in $γ$-Li$_2$IrO$_3$ are suggested as are torque measurements in other samples in which 
spin-liquids (Kitaev or not).

A state with $<\Omega> \ne 0$ is expected to have a quantized spin-Hall effect. The topological nature of such a state  was verified by Haldane and Arovas \cite{HA} by explicit calculation of a Chern number 2 (representing semion excitations) in a model of a hexagonal lattice with a effective Hamiltonian including the hermitian operator $S_i. (S_j \times S_k)$ supplemented with a Heisenberg Hamiltonian.  The connection to a quantized thermal Hall effect may follow, but this needs further investigation. 

It should be noted that while RuCl$_3$ may be considered two dimensional to a good approximation, $γ$-Li$_2$IrO$_3$, is three dimensional. While the two kinds of hexagonal planes (mentioned earlier) do not share any ions, we see no reason of symmetry that there is zero interactions between the magnetic ions in them.

Further theoretical work suggested is investigations of the effective Hamiltonians relevant for these compounds for the order parameter $<\Omega>$, conditions for it to have gap-less or gapped excitations, with and without an applied magnetic field, and the Chern class. Detailed investigations of torque  as a function of temperature and other techniques in the samples discussed above and those without the nuisance of an AFM order parameter are also suggested. Although topological, the chiral spin-order has a conventional $Z2 \times Z2$ symmetry. One would then expect it to occur as a phase transition (at a high temperature). Experiments to look for it should be done. A free-energy of the form of Eq. (\ref{BN}) has two branches which cross at $\theta = \pi/2, 3 \pi/2$. We have discussed the consequences for torque of always being in the lower energy equilibrium branch. One should however, in general, expect a hysteresis in the discontinuity of torque at the angles $\pi/2, 3\pi/2$. We suggest time-dependent experiments to look for it.
The chiral order parameter may also be around in other candidate "spin-liquids". It seem to us that torque measurements may be the most direct way to reveal them.\\

{\it Acknowledgements}: CMV acknowledges a conversation about the relation of the results of this work to the properties of the Kitaev model with Masaki Oshikawa, Lucille Savary and Senthil Todadri, and particularly wishes to thank Patrick Lee for several detailed discussions of the subject matter of this work.


\begin{thebibliography}{10}
\bibitem{footnote} 
Actually, RuCl$_3$ is slightly monoclinic. This is also seen in the data in the torsion data in Ref. \cite{Modic2}. In this paper, we ignore the monoclinicity in describing the results. This does not affect any of the conclusions drawn here
\end{thebibliography}
\end{document}